\documentclass[twocolumn,showpacs,preprintnumbers,amsmath,amssymb,superscriptaddress]{revtex4}

\usepackage{graphicx}
\usepackage{bm}
\usepackage{epsfig}

\begin{document}


\preprint{SFB/CPP-09-117, TTP09-44}
\title{Three-loop static potential}
\author{Alexander V. Smirnov}
\affiliation{Scientific Research Computing Center, Moscow State University, 119992 Moscow, Russia}

\author{Vladimir A. Smirnov}
\affiliation{Skobeltsyn Institute of Nuclear Physics of Moscow State University, 119992 Moscow, Russia}

\author{Matthias Steinhauser}
\affiliation{Institut f\"ur Theoretische Teilchenphysik, Karlsruhe
  Institute of Technology (KIT), D-76128 Karlsruhe, Germany}
\date{November 25, 2009}
\begin{abstract}
We compute the three-loop corrections to the potential of two heavy quarks.
In particular we consider in this Letter the purely gluonic contribution which
provides in combination with the fermion corrections of
Ref.~\cite{Smirnov:2008pn} the complete answer at three loops.
\end{abstract}


\maketitle


The potential between two heavy quarks constitutes a fundamental quantity in
Quantum Chromodynamics. It enters in a variety of physical processes like the
threshold production of top quark pairs and the description of charm and bottom
quark bound states. Furthermore, it is crucial for the understanding of
fundamental quantities of QCD, such as confinement. (See
Ref.~\cite{Brambilla:2004wf} for a recent review.)

The idea to describe a bound state of heavy coloured objects in analogy to the
well-established hydrogen atom, goes back to the middle of the
1970s~\cite{Appelquist:1974zd}. Shortly afterwards, about 30 years ago,
one-loop radiative corrections have been evaluated in the  
works~\cite{Fischler:1977yf,Billoire:1979ih}.
It took almost 20 years until the next order became
available~\cite{Peter:1996ig,Peter:1997me,Schroder:1998vy} which,
at that time, was a heroic enterprize.
The two-loop corrections turned out to be numerically quite important
which triggered several investigations to go beyond.
End of last year the fermionic corrections to the three-loop static potential
have been completed~\cite{Smirnov:2008tz,Smirnov:2008ay,Smirnov:2008pn}. In
this Letter  
we report about the pure gluonic part which completes the three-loop
corrections to the static potential.

We present our results for the static potential in momentum space where it
takes the form
\begin{eqnarray}
  \lefteqn{V(|{\vec q}\,|)=}
  \nonumber\\&&\mbox{}
  -{4\pi C_F\alpha_s(|{\vec q}\,|)\over{\vec q}\,^2}
  \Bigg[1+{\alpha_s(|{\vec q}\,|)\over 4\pi}a_1
    +\left({\alpha_s(|{\vec q}\,|)\over 4\pi}\right)^2a_2
    \nonumber\\&&\mbox{}
    +\left({\alpha_s(|{\vec q}\,|)\over 4\pi}\right)^3
    \left(a_3+ 8\pi^2 C_A^3\ln{\mu^2\over{\vec q}\,^2}\right)
    +\cdots\Bigg]\,.
  \label{eq::V}
\end{eqnarray}
Here, $C_A=N_c$ and $C_F=(N_c^2-1)/(2N_c)$
are the eigenvalues of the quadratic Casimir
operators of the adjoint and fundamental representations of the
$SU(N_c)$ colour gauge group, respectively, 
and $\alpha_s$ denotes the strong coupling in the $\overline{\rm MS}$ scheme.
The one- and two-loop coefficients 
$a_1$~\cite{Fischler:1977yf,Billoire:1979ih} 
and $a_2$~\cite{Peter:1996ig,Peter:1997me,Schroder:1998vy,Kniehl:2001ju} 
are given in Eq.~(4) of Ref.~\cite{Smirnov:2008pn} where also the
higher order terms in 
$\epsilon$, necessary for the three-loop calculation, are presented.
In Eq.~(\ref{eq::V}) we identify the renormalization scale $\mu^2$ and the
momentum transfer 
${\vec q}\,^2$. The complete dependence on $\mu$ can easily be restored with
the help of Eq.~(2) of Ref.~\cite{Smirnov:2008pn}.

A new feature of the three-loop corrections to $V(|{\vec q}\,|)$ is the
appearance of infrared divergences~\cite{Appelquist:1977es} which is
represented by the $\ln({\mu^2/{\vec q}\,^2})$ term in Eq.~(\ref{eq::V}).
It has been evaluated for the first time in
Refs.~\cite{Brambilla:1999qa,Kniehl:2002br} (see also 
Ref.~\cite{Brambilla:1999xf}); in Eq.~(\ref{eq::V}) we adopt the
$\overline{\rm MS}$ scheme which has been used 
in Ref.~\cite{Kniehl:2002br}.
Let us mention that the infrared divergence cancels in physical
quantities after including
the contribution where so-called ultrasoft gluons interact with the
heavy quark anti-quark bound state. An explicit result can, e.g., be found in
Ref.~\cite{Kniehl:2002br} where the cancellation has been
demonstrated in order to arrive at the measurable energy levels of
the heavy-quark system.
We note in passing that higher order logarithmic contributions to the
infrared behaviour of the static potential have been computed in
Refs.~\cite{Pineda:2000gza,Brambilla:2006wp}. 

\begin{figure}[t]
  \centering
  \leavevmode
  \epsfxsize=.42\textwidth
  \epsffile[160 330 560 490]{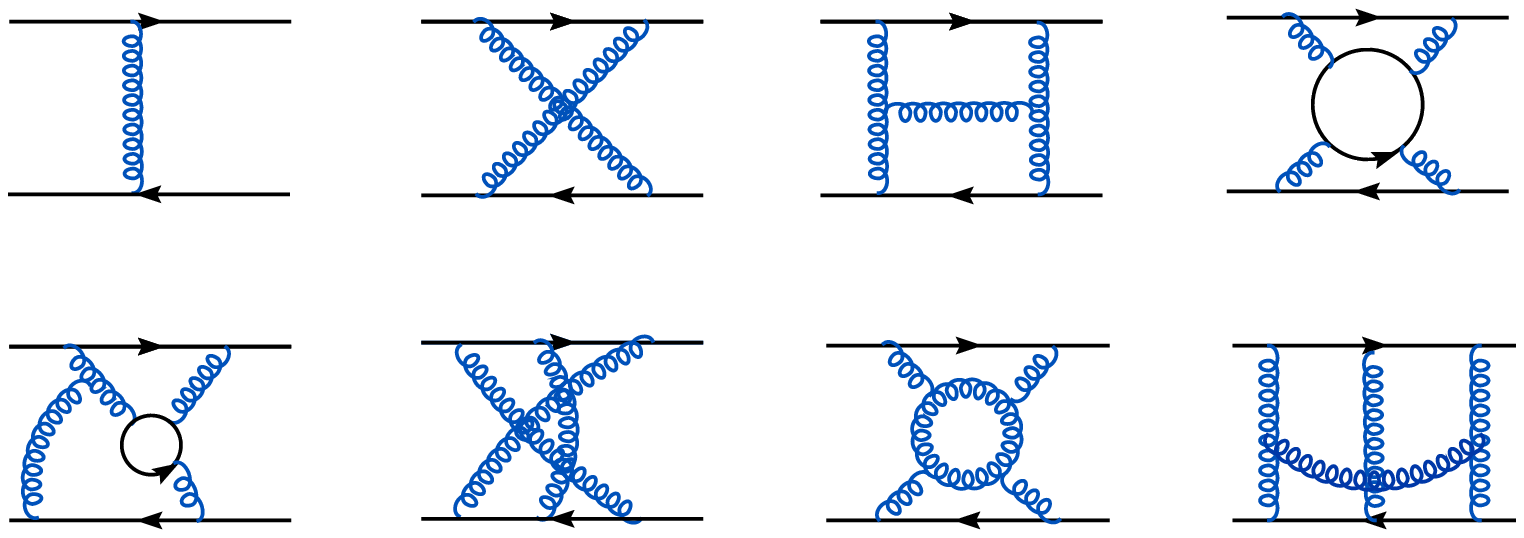}
  \caption{\label{fig::diags}Sample diagrams contributing to the
    static potential at tree-level, one-, two- and three-loop order.
    Solid and curly lines represent quarks and gluons, respectively.
    In the case of closed loops the quarks are massless; the external 
    quarks are heavy and treated in the static limit.
    }
\end{figure}

\begin{figure}[t]
  \centering
  \leavevmode
  \epsfxsize=.55\textwidth
  \epsffile[120 330 560 490]{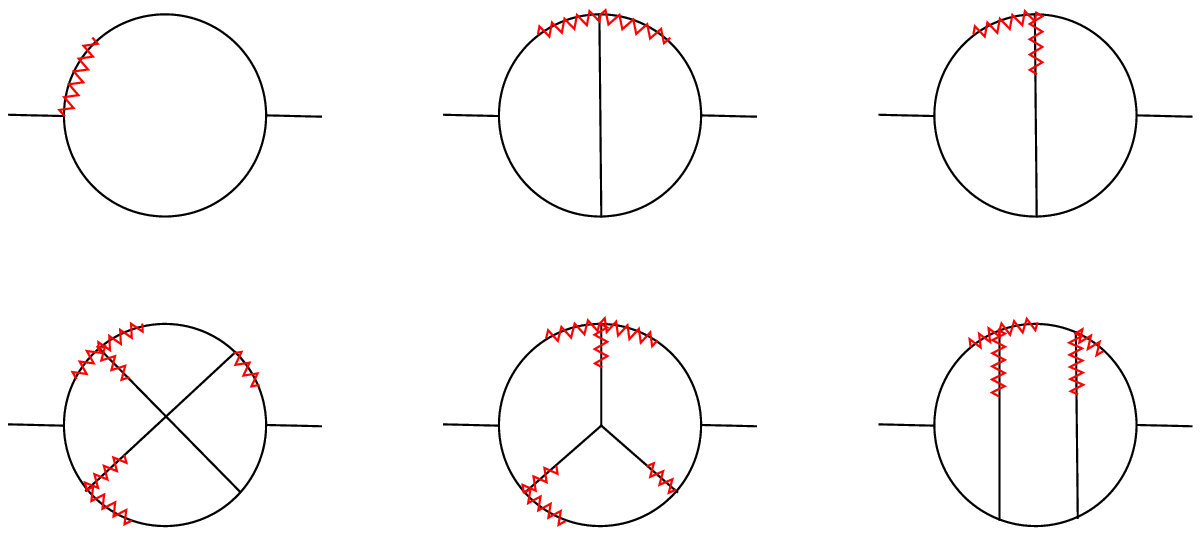}
  \caption{\label{fig::scalar}One-, two- and three-loop diagrams.
    The solid line stands for massless relativistic propagators and the 
    zigzag line represents static propagators.
    }
\end{figure}

\begin{figure}[t]
  \centering
  \leavevmode
  \epsfxsize=.50\textwidth
  \epsffile[100 330 560 420]{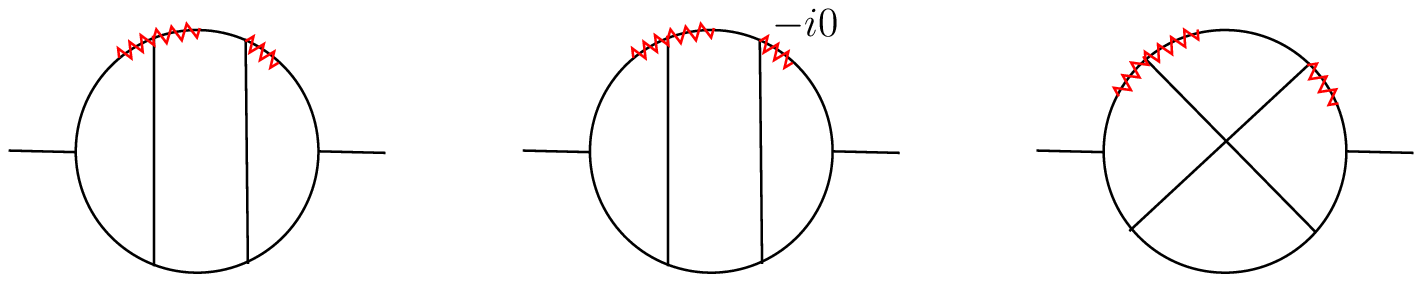}
  \caption{\label{fig::scalar2}Three-loop master integrals where the 
    ${\cal O}(\epsilon)$ part is only known numerically.
    The label ``$-i0$'' indicates that instead of the static propagator 
    $1/(p_0+i 0)$ there is the propagator $1/(p_0-i 0)$.
  }
\end{figure}

Before presenting our results for $a_3$ let us provide some technical details. 
We generate the four-point quark anti-quark amplitudes with the help of
{\tt QGRAF}~\cite{Nogueira:1991ex}.
Some sample diagrams up to three-loop order are shown in Fig.~\ref{fig::diags}.
In a next step they are processed further with {\tt q2e} and
{\tt exp}~\cite{Harlander:1997zb,Seidensticker:1999bb} where a mapping to the
diagrams of Fig.~\ref{fig::scalar} is achieved. 
The mapping to two-point functions is possible since
the only dimenionful quantity in our problem is given by the momentum 
transfer between the quark and the anti-quark. Although there is only one
mass scale in our problem technical complications arise from the 
simultaneous presence of static lines (zigzag lines) and
relativistic propagators (solid lines) which significantly increases
the complexity of the reduction to master integrals.
For this task we employ 
the program package {\tt FIRE}~\cite{FIRE} in order to achieve a reduction to
about 100 basic integrals, so-called master integrals. The latter have to be
evaluated in an expansion in $\epsilon$ which we achieve with
the help of the Mellin--Barnes method (see, e.g.,
Refs.~\cite{Smirnov:2004ym,Smirnov:2006ry,Czakon,Smirnov:2009up}). 
We managed to compute all the necessary
coefficients of the $\epsilon$ expansion of the master integrals
analytically with the exception of terms of order $\epsilon^1$ of the three
diagrams shown in Fig.~\ref{fig::scalar2}.
Results for the master integrals as well as more details
on their evaluation will be published elsewhere.
As a crucial tool providing very important numerical cross checks of the
analytical results we applied the program {\tt FIESTA}~\cite{FIESTA}
which is a convenient and efficient implementation of the sector
decomposition algorithm.
The colour factors of the individual Feynman diagrams have been computed with
program {\tt color}~\cite{vanRitbergen:1998pn}.

In our calculation we allowed for a general gauge parameter $\xi$ in the
gluon propagator. For individual diagrams we observe the appearance of
terms up to $\xi^6$. We have checked that the coefficients of the
$\xi^n$ ($n=1,\ldots,6$) terms are zero.

In order to present our results we decompose the three-loop
coefficient in the form
\begin{eqnarray}
  a_3 &=& a_3^{(3)} n_l^3 + a_3^{(2)} n_l^2 + a_3^{(1)} n_l + a_3^{(0)}
  \,,
\end{eqnarray}
where $n_l$ is the number of light quarks and the first three coefficients on
the right-hand side have been 
presented in Ref.~\cite{Smirnov:2008pn} (see Eq.~(6)).
Whereas for the fermionic contributions there are seven different
colour structures in the case of $a_3^{(0)}$ there are only two.
Note that the result of all colour structures containing a factor $C_F$ are
generated by iterations of lower-order contributions and thus do not
contribute to $a_3^{(0)}$. Diagrammatically such contributions are easily
identified since the corresponding Feynman integrals contain so-called pinch
contributions of the form $1/(p_0+i0)\times1/(p_0-i0)$ where $p_0$ is the
zeroth component of a loop momentum.
Our result for $a_3^{(0)}$ reads
\begin{eqnarray}
 a_3^{(0)} &=&
  502.24(1) \,\, C_A^3
  -136.39(12)\,\, \frac{d_F^{abcd}d_A^{abcd}}{N_A}
  \,.
  \label{eq::a3}
\end{eqnarray}
Similarly to the fermionic contribution new colour invariants
appear which can be traced back to Feynman diagrams as the third one
in the second row of Fig.~\ref{fig::diags}. 
Expressed in terms of $N_c$
one has $d_F^{abcd}d_A^{abcd}/N_A = (N_c^3 + 6N_c)/48$.
The coefficient of $d_F^{abcd}d_A^{abcd}$ has already been presented 
in Refs.~\cite{qcd09,radcor}, the coefficient of the $C_A^3$ term is new.

Let us now discuss the numerical effect of the three-loop contribution
to the static potential. Inserting the numerical results for the
coefficients $a_i$ in Eq.~(\ref{eq::V}) we obtain
\begin{eqnarray}
  V(|{\vec q}\,|)&=&-{4\pi C_F\alpha_s(|{\vec q}\,|)\over{\vec q}\,^2}
  \Bigg[1+\frac{\alpha_s}{\pi}\left(2.5833 - 0.2778 n_l\right)
    \nonumber\\&&\mbox{}
    +\left(\frac{\alpha_s}{\pi}\right)^2\left(28.5468 - 4.1471 n_l 
    + 0.0772 n_l^2 \right)
    \nonumber\\&&\mbox{}
    +\left(\frac{\alpha_s}{\pi}\right)^3\left(
    209.884(1) -51.4048 n_l 
    \right.\nonumber\\&&\left.\mbox{}
    + 2.9061 n_l^2
    - 0.0214 n_l^3\right)
    +\cdots\Bigg]\,,
  \label{eq::Vnum}
\end{eqnarray}
where $\mu^2={\vec q}\,^2$ has been adopted in order to suppress the 
infrared logarithm and the 
ellipses denote higher order terms in $\alpha_s$.
The term ``209'' in the three-loop coefficient 
receives a large contribution (``211'') from the $C_A^3$ term whereas
the new colour structure only contributes with a coefficient ``$-2$''.
From Eq.~(\ref{eq::Vnum}) we observe at one-, two- and three-loop
order a large screening of the non-fermionic contributions by the $n_l$ terms
which is
most prominent in the case of $a_3$ for $n_l=5$. Here the difference
between $a_3^{(0)}$ and the fermionic contribution is one order
smaller than the individual pieces.

In Tab.~\ref{tab::a123} we show the numerical evaluation of the square bracket
of  
Eq.~(\ref{eq::Vnum}) for the charm, bottom and top quark case, i.e. for
$n_l=3,4$ and~5, adopting the appropriate values of $\alpha_s$.
For charm the three-loop corrections are almost as big as the one- and
two-loop contributions whereas for bottom the three-loop contribution
is already a factor of four smaller than the two-loop one.
In the case of the top quark one observes a good convergence: the
three-loop term is already a factor ten smaller than the two-loop
counterpart. 

\begin{table}[t]
  \begin{center}
    \begin{tabular}{c|l|l|l|l}
      $n_l$ & $\alpha_s^{(n_l)}$ & 1 loop & 2 loop & 3 loop \\
      \hline
      3 & 0.40 & 0.2228 & 0.2723 
      & 0.1677 \\
      4 & 0.25 & 0.1172 & 0.08354 
      & 0.02489\\
      5 & 0.15 & 0.05703 & 0.02220 
      & 0.002485
    \end{tabular}
    \caption{\label{tab::a123}Radiative corrections to the potential
      $V(|{\vec q}\,|)$ where the tree-level result is normalized to 1
      (cf. Eq.~(\ref{eq::Vnum})).
      In the second column we also provide the numerical value of
      $\alpha_s$ corresponding to the soft scale where $\mu\approx m_q\alpha_s$
      ($m_q$ being the heavy quark mass).}
  \end{center}
\end{table}

As already mentioned above, $V(|{\vec q}\,|)$ itself is not
a physical quantity. Hence let us consider the ground state energy $E_1$
of a heavy
quarkonium system which has been evaluated to the third order in perturbation
theory in Ref.~\cite{Penin:2002zv} where the contribution
from $a_3$ has been kept unevaluated. We are now in the position to complete
the numerical analysis. It is convenient to write the perturbative
contribution to $E_1$ in the form 
\begin{eqnarray}
  E_1^{\rm p.t.} &=& E_1^C + \delta E_1^{(1)} +
  \delta E_1^{(2)} +\delta E_1^{(3)} + \ldots\,,
\end{eqnarray}
with the Coulomb energy $E_1^C=-C_F^2\alpha_s^2 m_q/4$. $m_q$ is the heavy
quark mass and the superscript in brackets indicates the order in perturbation
theory. Adopting for the renormalization scale the choice 
$\mu_S=C_F\alpha_s(\mu_S)m_q$ we obtain 
\begin{eqnarray}
  \delta E_1^{(3)}\Big|_{\rm charm} \!\!\!\!\!&=&
  \alpha_s^3 E_1^C \!\left( 129.79 + 5.241\Big|_{a_3} \!\!\!
    + 15.297 \ln(\alpha_s) \right)
  ,\nonumber\\
  \delta E_1^{(3)}\Big|_{\rm bottom} \!\!\!\!\!\!\!&=&
  \alpha_s^3 E_1^C \!\left( 104.82 + 3.186\Big|_{a_3} \!\!\!
    + 15.297 \ln(\alpha_s) \right)
  ,\nonumber\\
  \delta E_1^{(3)}\Big|_{\rm top} \!\!\!\!\!&=&
  \alpha_s^3 E_1^C \!\left( 83.386 + 1.473\Big|_{a_3} \!\!\!
    + 15.297 \ln(\alpha_s) \right)
  ,\nonumber
  \\
\end{eqnarray}
where the contribution from $a_3$ has been marked separately.
One observes that the numerical effect amounts between 1 and 4\% of the
non-logarithmic constant.

Finally, it is interesting to compare our results with the predictions
obtained on the basis of certain assumptions on the perturbative
expansion. In Ref.~\cite{Chishtie:2001mf} a Pad\'e approximation in
the coupling constant has been performed whereas the findings of 
Ref.~\cite{Pineda:2001zq} are based on renormalon studies.
For $a_3^{(0)}/4^3$ they predict 313 and 292, respectively, which
overshoots the exact result by 40 to 50\%.

More recently, a detailed comparison of the perturbative result with
lattice simulations has been performed~\cite{Brambilla:2009bi}
with the aim to extract $a_3^{(0)}$. After transforming the result
of Ref.~\cite{Brambilla:2009bi} to momentum space using the formulae
provided in their Appendix one obtains $202 \le a_3^{(0)}/4^3 \le
337$. Thus the lower limit of the (relatively big) interval covers the exact
result. 

To conclude, in this Letter the three-loop corrections to the static
potential have been completed by evaluating the gluonic contribution. 
Our main result can be found in Eq.~(\ref{eq::a3}) where the
three-loop coefficients are given for general colour structure.
Numerical sizeable corrections are observed for the non-fermionic
contributions which are partly canceled by the fermionic
corrections evaluated in Ref.~\cite{Smirnov:2008pn}.

Let us stress that the static potential constitutes a fundamental quantity of
QCD. It represents a building block in many
physical quantities like the determination of the 
bottom quark mass from the $\Upsilon(1S)$ bound state
or the third-order correction to top quark threshold production cross section
at a future electron positron linear collider which would result in the most
precise 
value for the top quark mass.
The static energy is also a crucial object when comparing
perturbation theory and lattice simulations (see, e.g.,
Refs.~\cite{Bali:1999ai,Necco:2001gh,Pineda:2002se,Brambilla:2009bi}). 
We also want to mention the extraction of the strong coupling constant 
from lattice simulations where again the static potential and in
particular $a_3$ plays an important
role~\cite{Davies:2008sw,Maltman:2008bx}.

{\it Note added}: While finishing this paper we became aware of the preprint
C.~Anzai, Y.~Kiyo and Y.~Sumino, ``Static QCD potential at three-loop order,''
arXiv:0911.4335 [hep-ph], where also $a_3^{(0)}$ has been computed. We agree
with their Eq.~(10), however, obtain a better precision.



Acknowledgements. 
We would like to thank Alexander Penin for many useful discussions and
communications. We are grateful to Alexey Pak for assistance and to Johann
K\"uhn for useful comments.
This work is supported by DFG through SFB/TR~9
``Computational Particle Physics'' and RFBR, grant 08-02-01451.




\end{document}